\documentclass[twocolumn,preprintnumbers,
endnote,prl]{revtex4-1}
\usepackage{graphicx}

\usepackage{color}
\usepackage{subfig}
\usepackage{amsmath}
\usepackage{cancel}

\newcommand{\vev}[1]{\langle {#1} \rangle}
\newcommand{\lsim}{\lesssim}
\newcommand{\gsim}{\gtrsim}

\newcommand{\ord}[1]{\mathcal{O}{(#1)}}
\newcommand{\beq}{\begin{equation}}
\newcommand{\eeq}{\end{equation}}
\newcommand{\bea}{\begin{eqnarray}}
\newcommand{\eea}{\end{eqnarray}}

\newcommand{\mP}{M_{\rm P}}
\newcommand{\rmP}{\bar{M}_{\rm P}}
\newcommand{\Neff}{N_{\rm eff}}
\newcommand{\tsm}{T_{\rm SM}}

\begin{document}

\pagestyle{plain}

\title{\boldmath Fuzzy Dark Matter from Infrared Confining Dynamics}

\author{Hooman Davoudiasl
\footnote{email: hooman@bnl.gov}
}

\author{Christopher W. Murphy
\footnote{email: cmurphy@bnl.gov}
}

\affiliation{Department of Physics, Brookhaven National Laboratory,
Upton, NY 11973, USA}


\begin{abstract}

A very light boson of mass $\ord{10^{-22}}$~eV may potentially be a viable dark matter (DM) candidate 
which can avoid phenomenological problems associated with cold DM.  Such ``fuzzy DM (FDM)" may naturally 
be an axion with a decay constant $f_a \sim 10^{16} \div 10^{18}$~GeV, and a mass $m_a \sim \mu^2/f_a$ with $\mu\sim 10^2$~eV.  Here we propose a concrete model where $\mu$ arises as a dynamical scale
from infrared confining dynamics, analogous to QCD. Our model is an alternative to the usual approach of generating $\mu$ through string theoretic instanton effects.  
We outline the features of this scenario that result from  
various cosmological constraints.  We find that those constraints are suggestive of a period of mild of inflation, perhaps from a 
strong first order phase transition, that reheats the Standard Model (SM) sector only.   A typical prediction 
of our scenario, broadly speaking, is a larger effective number of neutrinos compared to  the SM value $\Neff \approx 3$, as 
inferred from precision measurements of the cosmic microwave background.  Some of the new degrees of freedom may be 
identified as ``sterile neutrinos," which may be required to explain certain neutrino oscillation anomalies. Hence, aspects of our scenario could be testable in terrestrial experiments, which is a novelty of our FDM model.

\end{abstract} 
\maketitle

The dark matter (DM) riddle -- the question of what composes nearly 85\% of all matter in the Universe -- has yet to find a 
resolution.  What is known about this mysterious substance, which has no viable candidate in the Standard Model (SM), is only gleaned from its gravitational effects on cosmology and 
astrophysics.  Generally speaking, DM seems to interact only feebly if at all with any form of matter, dark or otherwise.  

Various models of new physics have been considered for DM, often also 
addressing other open questions of particle physics.  Many such models lead to cold DM (CDM), a particle often characterized by 
mass scales $m\gsim$~GeV.  An interesting alternative is axion DM, which initially arose in the 
extensions of the Peccei-Quinn (PQ) resolution of the strong charge conjugation parity symmetry (CP) 
problem~\cite{Peccei:1977hh, Wilczek:1977pj, Weinberg:1977ma} (related to why the CP violating parameter in QCD seems to be  $\lsim 10^{-9}$~\cite{Olive:2016xmw}).  

If the decay constant for the PQ axion, $f_a$, is of order $10^{12}$~GeV, it can also 
be a good CDM candidate, addressing two problems at once~\cite{Preskill:1982cy, Abbott:1982af, Dine:1982ah}.  Yet, one is faced with the 
question of why such an anomalous symmetry should exist in Nature.  Fortunately, string theory, a promising 
framework for quantum gravity, seems to offer a multitude of axion candidates, see for example Ref.~\cite{Svrcek:2006yi}.  However, 
the decay constant of axions in these models are typically $f_a \sim 10^{16} \div 10^{18}$~GeV, between the typical scale 
of grand unified theories and the reduced Planck mass $\rmP \approx 2.4 \times 10^{18}$~GeV.  

The above string theory predictions for $f_a$, without 
severe tuning of initial conditions, would yield excessive amounts of DM, in serious conflict with cosmological observations.  This 
situation can be changed if the potential for the axion is not assumed to be generated by QCD, and instead non-perturbative mechanisms from string theory itself are employed.  Here, various instanton effects may be used to generate a small 
scale, $\mu \sim 10^2$~eV, for the axion potential from underlying high scales of the theory associated with gravity and supersymmetry breaking~\cite{Svrcek:2006yi}.  Indeed, this fact can be used to entertain the intriguing possibility that DM is made of ultra light bosons of mass $\sim 10^{-22}$~eV, with a De Broglie wave length of $\ord{\rm kpc}$~\cite{Hui:2016ltb}. Such DM could address the shortcomings of the CDM scenario, 
typically related to its implications for small scale galactic structure and dynamics.  This type of DM has been dubbed ``fuzzy DM (FDM)"~\cite{Hu:2000ke} and can be naturally identified with an almost massless axion, see for instance~\cite{Press:1989id, Sin:1992bg, Peebles:2000yy, Goodman:2000tg, Amendola:2005ad, Viel:2013apy, Schive:2014dra, Hlozek:2014lca, Kim:2015yna, Marsh:2015xka}.  If the scale $\mu\sim 100$~eV is generated from ultraviolet scales, its manifestation does not 
entail any dynamics in the low energy theory.  

Alternatively, one may also assume that the 
above scale $\mu$ is related to non-trivial dynamics at very low energies, 
in analogy with QCD (as alluded to in~\cite{Hui:2016ltb}).  
In this work, we adopt this 
possibility and pioneer the study of its typical phenomenology.  To the best of our knowledge,  this is the first time a concrete model of this type has been proposed for FDM. Given the hierarchy of scales 
of $\ord{10^6}$ between this new dynamics and 
QCD, we refer to it as micro-QCD ($\mu$-QCD for short) whose confinement scale is denoted by $\mu$.  For simplicity and 
in order to take advantage of the well-known QCD features, we will choose $SU(3)_\mu$ as the gauge group and further assume 
that there are ``$\mu$-quarks'' that get masses from a low energy, $\ord{\rm keV}$, Higgs mechanism.  As a toy analogue of the 
SM case, we will also assume that there are new light ``colorless" fermions that the $\mu$-hadrons can decay into; these ``$\mu$-leptons'' may be identified as ``sterile" neutrinos. We emphasize that -- in comparison to the string theoretic generation of the scale $\mu$ -- our model entails not only extra cosmological signatures, but also the novel possibility of terrestrial probes at neutrino oscillation experiments.

The current constraints imposed by successful big bang nucleosynthesis (BBN) suggest that no more than one 
new degree of freedom may be present aside from the SM content at temperatures of $\ord{\rm MeV}$.  Given that 
we would like to introduce new gauge dynamics with a confining scale of $\ord{\rm 10^2 eV}$, we need to suppress the effect of 
those new states on the evolution of the Universe during BBN.  Similar constraints are also present from precision 
measurements of the cosmic microwave background (CMB).  These constraints can be largely avoided if 
the new degrees of freedom do not contribute more than $\Delta N_{\rm eff} \sim 0.1$.  
This can be achieved if the temperature of the $\mu$-sector $T'$ is  
is sufficiently smaller than that of the SM sector $\tsm$. By ``SM-sector" we mean all 
the states that are in the SM or its extensions that decouple from the new infrared particles, which we will refer to as the $\mu$-sector. As we will see, $\ord{30}$ new degrees of freedom need to be added, and at low energies would make a sufficiently small contribution if $T'\lsim \tsm/4$. 

The cooler temperature of the $\mu$-QCD sector may be due to its decoupling from the SM sector at some high scale, 
with subsequent transfer of entropy to the SM sector only.  This is similar to the way background neutrinos are assumed to be 
colder than photons for $\tsm \lsim 1$~MeV, in standard cosmology.  After the two sectors decouple, decoupling of the 
SM sector states would only increase $\tsm$ and not $T'$.  To have $T'\lsim \tsm/4$, the 
number of the SM sector states that leave the thermal bath prior to BBN must be $\gsim 10 \times 4^3$, where we have approximated the relativistic degrees of freedom during BBN as $\sim 10$.   

There is another possibility that 
could provide a sufficiently small $T'$, which is a period of mild inflation~\cite{Davoudiasl:2015vba} (see also Ref.~\cite{Carlson:1992fn}).  
Such a possibility may be realized through thermal effects (``thermal inflation"~\cite{Lyth:1995ka}) or else as a consequence of a strong first 
order phase transition that briefly lingers in a false vacuum before tunneling.  In fact, such a process may be a natural 
component of a baryogenesis mechanism, providing the requisite Sakharov condition of departure from equilibrium.  The required number of $e$-foldings is $\gsim 1$, which will universally cool all sectors.  As long as the consequent reheating 
involves only the SM sector and its possible extensions, the $\mu$-sector would remain at the cooler temperature.  However, 
further decoupling, as in after $\mu$-QCD confinement, will reheat the $\mu$-sector somewhat, which will be considered in our discussion later. 

We will assume an $SU(3)_\mu$ 
interaction and at least one flavor of $\mu$-quarks, denoted by $\psi$, in the fundamental representation.  We will 
further assume a $Z_2$ parity under which $\psi$ has a chiral representation, with  $Z_2(\psi_L) = +1$ and $Z_2(\psi_R) = -1$.  
In order to have a non-trivial axion potential for the FDM, we cannot have massless quarks.  We hence introduce 
a real scalar $\phi$ with $Z_2(\phi) = -1$, whose vacuum 
expectation value (vev) breaks the $Z_2$ parity and endows $\mu$-quarks 
with mass through the interaction $y_\psi \phi\, \bar \psi_L \psi_R + \text{\small H.C.}$, with $\vev{\phi} \neq 0$.

Assuming that the lightest mesons are lighter than the baryons, the baryons and anti-baryons will annihilate into those mesons. The mesons are expected to have masses $\gsim 100$~eV 
in our scenario, and 
may come to dominate the energy density before the usual matter-radiation equality at $\tsm \sim 1$~eV \footnote{For example, stable hadrons associated with a new confining dynamics could themselves be dark matter candidates, which would call into question the need for an axion in the first place. See Ref.~\cite{Francis:2016bzf} for an example of such a scenario with $N_{\mu} = 2$ and $N_f = 1$.}.  To avoid unwanted effects we will assume that the lightest $\mu$-hadrons can annihilate quickly 
into light, colorless fermions $n$ which get masses of $\lsim 1$~eV 
from their coupling to $\phi$ given by $y_n \phi \,\bar n_L n_R + \text{\small H.C.}$, where $Z_2(n_L) = +1$ and 
$Z_2(n_R)  = -1$. 

In close analogy with the QCD axion, non-perturbative effects set up a potential for the FDM axion $a$.  
Following the well-known case of QCD axion, the energy density of the FDM axion zero mode 
is given by~\cite{Preskill:1982cy, Abbott:1982af, Dine:1982ah} 
\beq
\rho_a = \frac{1}{2} m_{a, i} m_{a, f} A_i^2  \left(\frac{g_{s, f}}{g_{s, i}}\right) \left(\frac{T_f}{T_i}\right)^3\,,
\label{rhoa}
\eeq 
where $m_a$ is the axion mass, the subscripts $i$ and $f$ refer to initial and final values.  In the above, $A_i$ is the 
initial amplitude of oscillations, $g_s$ denotes the relativistic 
degrees of freedom in equilibrium in the SM sector, and $T$ is its corresponding temperature: $T\equiv 
\tsm$.  Here, $T_i$ is the temperature at which the axion FDM starts to oscillate, given by 
\beq
m_{a, i} \approx 3 H (T_i)\,,
\label{mai}
\eeq
where $H(T_i) = 1.66 g_i^{1/2} T_i^2/\mP$ is the Hubble scale at $T=T_i$ and $\mP\approx 1.2 \times 10^{19}$~GeV. Since we are interested in the regime $T\ll 1$~MeV, we will assume $g_{s,i}=g_{s,f}=g_s$.  
We also have $T_f \approx 2.3 \times 10^{-4}$~eV, corresponding to the CMB temperature today.

The mass $m_a$ as a function of temperature $T'$ in the $\mu$-sector can be written as~\cite{Gross:1980br, Davoudiasl:2006bt} 
\bea\label{maT}
m_a(T') &\approx& \xi^{\frac{N_f}{2}} \kappa^{N_{\mu}} \left(\frac{c_N}{2 \eta}\right)^{\frac{1}{2}} 
\left(\frac{m_\psi}{\mu}\right)^{\frac{N_f}{2}} \left(\frac{\mu}{\pi T'}\right)^{\eta}\\ \nonumber
&\times&\left(\frac{\mu^2}{f_a}\right)
\ln^{N_{\mu}}(\pi T'/\mu)\,,
\eea
where $\xi \approx1.34$, $N_{\mu}$ is the number of $\mu$-colors, which we will set at $N_{\mu}=3$ 
for the rest of our discussion, and $N_f$ is the number of $\mu$-quark flavors.  We also have 
$c_N\approx 0.26/[\xi^{N_{\mu}-2}(N_{\mu}-1)!(N_{\mu}-2)!]$, $\kappa \equiv (11N_{\mu}-2 N_f)/6$, and 
$\eta \equiv \kappa + N_f/2 -2$. The finite-temperature mass, $m_a(T^{\prime})$, should be a monotonically decreasing function of $T^{\prime}$. Eq.~\eqref{maT} is the leading high temperature approximation to $m_a(T^{\prime})$, and as such it only gives sensible results for $T^{\prime} \gg \mu / \pi$, a point which we will return to later.

To go further we will choose a minimal matter content with $N_f=1$.  (See Ref.~\cite{Creutz:2006ts} for a review of $N_f=1$ QCD.) In this case, due to a $U(1)$ anomaly, 
no chiral symmetries of the $\mu$-QCD will survive and there are no light Goldstone boson counterparts to pions in the SM. 
By analogy with the mass of $\eta'$ in the SM, we may expect that the mass 
of the lightest meson, denoted here by $\Pi$, is given by $m_{\Pi} \approx 4 \mu$.  We will assume that there is 
only one flavor of $n$, for simplicity.  

In our scenario, $\mu$-hadrons are assumed to annihilate into a $\Pi$ meson 
population, which will then in turn annihilate via $\Pi \Pi \to n \bar{n}$ or else decay, $\Pi \to n \bar{n}$, promptly.  It is interesting that 
the parameters of the FDM axion allow for 
$\Pi \to \bar n n$ to be efficient, as we will explain below.   

Curiously, for a FDM axion of mass $m_a \sim 10^{-22}$~eV, the period of oscillation is $\sim 1$ year (see also Refs.~\cite{Stadnik:2015kia, Stadnik:2016zkf,Blas:2016ddr}.  Hence, 
if the lifetime of $\Pi$ is short compared to this time scale, the $\Pi$ population 
can decay into $n \bar n$ while the magnitude of 
the effective CP violating angle $\theta_\mu^{eff} \sim a/f_a \sim 1$.  We will later show that this is 
the case in our typical parameter space.

The total number of degrees of freedom in the $\mu$-sector, given the above minimal assumptions, is 
$N' = (4\times 3 + 4) \times 7/8 + 16 + 1 = 31$.  Hence, as a reasonable value, we may choose 
$T'=T/4$, which would yield $\Delta N_{\rm eff}\approx 0.12$, at temperatures $\gsim \mu$.  This is 
sufficiently small during BBN, however we should also make sure that at $T\sim 1$~eV, we do not end up with 
an unacceptable $\Delta N_{\rm eff}$.  Assuming that the 31 degrees of freedom annihilate into $n$ and $\bar n$, 
we would end up with $N'=3.5$ and $T' = (31/3.5)^{1/3} T/4 \approx 0.5 \,T$ which yields $\Delta N_{\rm eff} \approx 0.25$, 
which is a reasonable value given current uncertainties~\cite{Ade:2015xua}.

To obtain the contribution of FDM to the cosmic energy budget, we need to know the $T=0$ value of the axion mass, here denoted by $m_a$.  One can show that~\cite{Georgi:1986df} 
\beq
m_a^2 \approx - \frac{\langle 0| \bar \psi \psi |0\rangle}{f_a^2 \,\text{Tr}(M^{-1})}\,,
\label{ma}
\eeq
where the $\mu$-quark condensate $\langle 0| \bar \psi \psi |0\rangle \approx - \mu^3$ and $M$ is the diagonal 
$\mu$-quark mass matrix.  We will typically set $m_a = 10^{-22}$~eV in the rest of our analysis, as motivated by the 
FDM scenario. Note that the above equation would then yield a value for $\mu$ given a choice for $M$.  
 
\begin{figure}
  \centering
\includegraphics[width=0.45\textwidth]{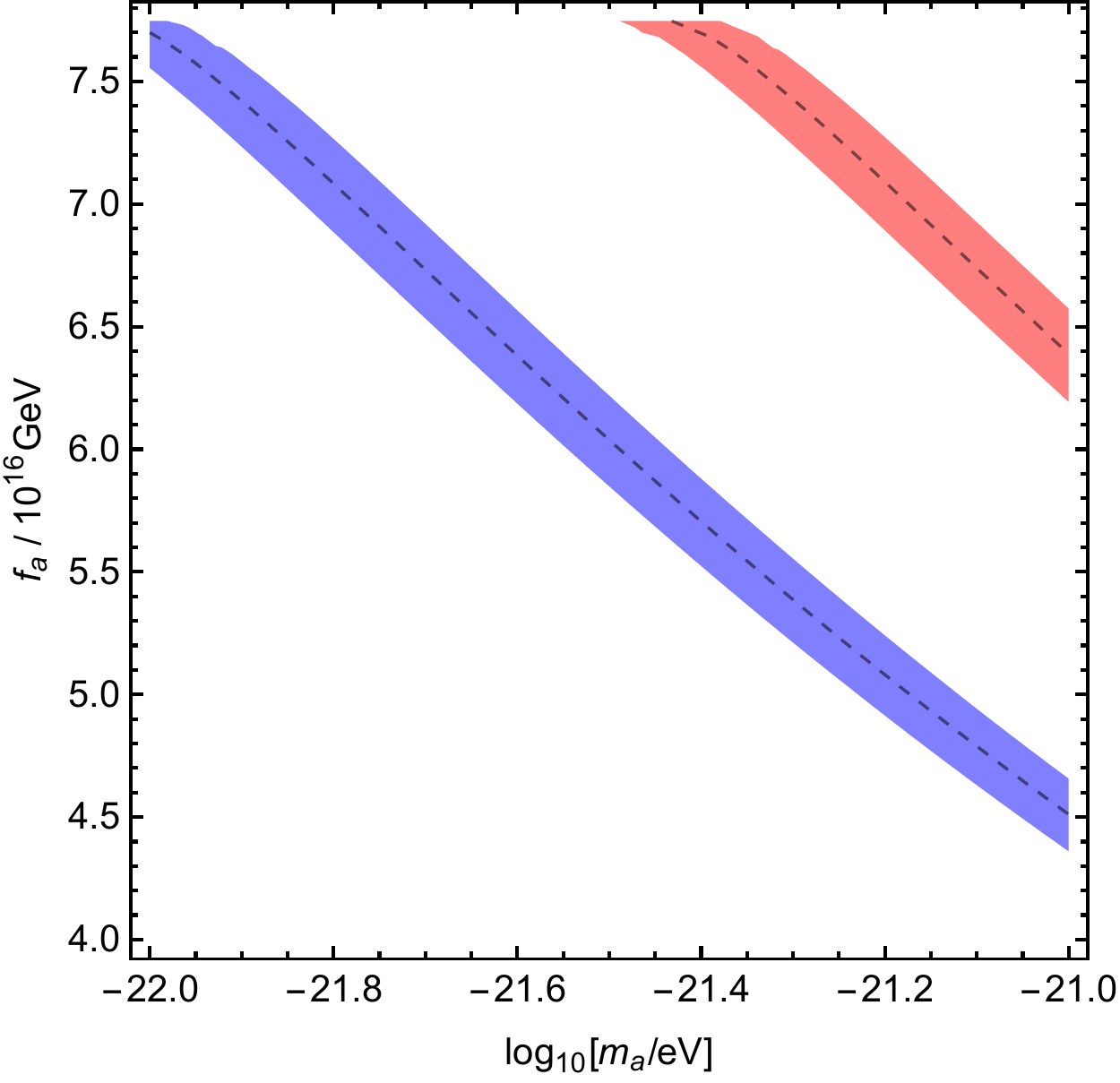}
   \caption{FDM density in the $f_a$ vs. $m_a$ plane with $N_f=1$, $T'=T/4$, and $m_\psi = 0.5 \mu$. The blue (bottom) and red (top) bands reproduce the correct value of $\Omega_{\text{DM}}$ at the 95\%~CL for $A_i = f_a \text{ and } f_a / \sqrt{2}$, respectively, with the dashed lines indicating the central value.}
 \label{fig:FDM}
 \end{figure}

Cosmological observations yield $\Omega_{\rm DM} \approx 0.258 \pm 0.008$, where $\Omega_{\rm DM}\equiv \rho_a/\rho_c$ 
and $\rho_c \approx (2.6 \times 10^{-3}~\text{eV})^4$ is the critical energy density~\cite{Ade:2015xua}.   In the regime of interest, $T\ll 1$~MeV, 
we have $g_i \approx 3.36$ (only considering the SM sector).  With $N_f=1$, $T'=T/4$, $m_\psi = 0.5 \mu$, and using Eqs.~(\ref{rhoa}), (\ref{mai}),~(\ref{maT}), and (\ref{ma}), 
we find the observed value of $\Omega_{\rm DM}$ can be reproduced. This is illustrated in Fig.~\ref{fig:FDM}, which shows the FDM density in the $f_a$ vs. $m_a$ plane. The blue (bottom) and red (top) bands indicate the correct value of $\Omega_{\text{DM}}$ at the 95\%~CL for $A_i = f_a \text{ and } f_a / \sqrt{2}$, respectively, with the dashed lines indicating the central value.  

Note that as $T'\to 0$, the expression in Eq.~\eqref{maT}, which corresponds to the leading high temperature approximation, does not yield the expected zero temperature value of $m_a$.  Hence, we only consider solutions to Eq.~\eqref{mai} where $T^{\prime} > 2.27 \mu / \pi$, which corresponds to the temperature at which the right-hand side of~\eqref{maT} is maximal.   

The presence of new, neutral degrees of freedom with masses of order eV is suggestive of a connection with neutrino physics. In fact, there have been a number of measurements over the years that have challenged the three neutrino paradigm. These include the short baseline experiments LSND~\cite{Athanassopoulos:1995iw, Aguilar:2001ty} and MiniBooNE~\cite{Aguilar-Arevalo:2013pmq}, which reported evidence for $\bar{\nu}_{\mu} \to \bar{\nu}_{e}$ oscillations outside the realm of explanation in a three flavor neutrino model. In addition, the so-called reactor~\cite{Mention:2011rk} and gallium~\cite{Abdurashitov:2005tb, Giunti:2012tn} anomalies have a natural interpretation in terms of ``sterile'' neutrinos. The preferred parameter space is $m_n \sim 1$~eV with a mixing angle $\theta \sim 0.1$. These parameters are used as benchmark points in searches for sterile neutrinos in other experiments including ICARUS~\cite{Antonello:2013gut} and IceCube~\cite{TheIceCube:2016oqi}. 

A major model building challenge associated with ``sterile'' neutrinos with a sizable mixing angle is explaining the precise measurement of $N_{\text{eff}} = 3.15 \pm 0.23$ by the Planck collaboration~\cite{Ade:2015xua}. The most common proposal to solve this problem is that the ``sterile'' neutrinos actually have interactions of some sort~\cite{Hannestad:2013ana, Dasgupta:2013zpn, Bringmann:2013vra, Archidiacono:2014nda, Saviano:2014esa, Mirizzi:2014ama, Chu:2015ipa, Cherry:2016jol}. The possibility of a late phase transition has been explored as well~\cite{Vecchi:2016lty}.

Our mechanism for evading this bound proceeds as follows. We introduce a second scalar field with an approximate shift symmetry $S \to S + S_0$ that makes all of interactions with other particles very weak, and some new right-handed neutrinos, $N_R$.\footnote{Ref.~\cite{Berlin:2016woy} considers the case where an ultra-light scalar field, such as an FDM axion, mediates the neutrino interactions.} Furthermore, we assign $S$ and $N_R$ odd parity under a new $Z_2$ symmetry under which all the other particles are even. The relevant interactions are 
\begin{align}
\label{eq:Vphi}
\mathcal{L} &\supset - \lambda_{\phi} \left(\phi^2 - v_{\phi}^2\right)^2 + \lambda \phi^2 S^2 - \lambda_S S^4 \\
&- g S \bar{n}_L N_R - \frac{1}{M} S \tilde{H} \bar{L}_L N_R + \text{\small H.C.} , \nonumber
\end{align}
with $\langle \phi \rangle \equiv v_{\phi}$, and where $g$, $\lambda_{\phi}$, $\lambda$, $\lambda_S$, and $M$ are parameters to be determined. The classical stability of the scalar potential requires 
\begin{equation}
\label{eq:stab}
\lambda_{\phi} \lambda_S \gsim \lambda^2 . 
\end{equation}
In addition to the previously introduced $\phi$, $n_L$, and $N_R$, we see that $S$ interacts with the active SM neutrinos contained in $L_L = (\nu_L, \ell_L)^T$ and the SM Higgs boson, $\tilde{H} = \epsilon_{ij} H^{j \dagger}$. For simplicity we take $M = f_a$ as the FDM axion decay constant is sufficiently high to suppress the active-sterile neutrino interactions. We assume the renormalizable interactions of $S$ with $H$ as well as a hard mass term for the $S$ are small enough that they can be ignored.  

The idea is that the active neutrinos do not have masses and do not mix with the sterile neutrinos until $S$ gets a vev, and $S$ does not get a vev until temperature of the universe cools such that $\mu$-Higgs can settle down into its vev. Demanding no mixing until after the active neutrinos decouple from the electron-photon bath places the bound $v_{\phi} \lsim 100$~keV.  Once the temperature of the universe cools sufficiently, i.e. below 100~keV, $\phi$ gets a vev, which then leads to a vev $\langle S \rangle \equiv v_S \sim \sqrt{\lambda / \lambda_S} v_{\phi}$. 

The vev of $S$, once it turns on after evading the bound from $N_{\text{eff}}$, gives rise to the masses of the active neutrinos, and the mixing between the active and sterile neutrinos. Since $\vev{H} / f_a \sim 10^{-14}$, to achieve the correct value for the active neutrino mass, $m_{\nu} \sim 0.1$~eV, we need $v_S \sim 10$~TeV. This leads to the following relation $\lambda_S \sim 10^{-16} \lambda$, in which case~\eqref{eq:stab} implies $\lambda_{\phi} \gsim 10^{16} \lambda$. The correct mixing angle, $\theta \sim 0.1$, is then obtained for $g \sim 10^{-13}$, consistent with the assumption of an approximate shift symmetry. If $m_{\phi}$ is light enough the population of $\Pi$ can be depleted through the $\phi$ mediated process $\Pi \Pi \to n \bar{n}$. Assuming $T \lsim m_{\Pi} \ll m_{\phi}$ we find roughly 
\begin{equation}
\Gamma (\Pi \Pi \to n \bar{n}) \sim T'^3 \sigma \sim T'^3 \frac{1}{64 \pi m_{\Pi}^2} \left(\frac{m_{\psi} \mu m_n m_{\Pi}}{v_{\phi}^2 m_{\phi}^2}\right)^2 ,
\end{equation}
where $(64 \pi m_{\Pi}^2)^{-1}$ is our estimate for the phase space of the cross section.  Here, we have also assumed $m_\psi \sim 100$~eV, $m_n\sim 1$~eV, and $v_\phi \sim 100$~keV.  This process stays efficient well below $T \sim 100$~eV, for $\lambda_{\phi} \lsim 10^{-3}$, which implies $\lambda \lsim 10^{-19}$ and $\lambda_S \lsim 10^{-35}$.  These values are also consistent with a shift symmetry for $S$.  With these choices of parameters, $m_\phi \lsim \text{few}$~keV.  These parameters lead to a small mass for $S$, $m_S \sim \sqrt{\lambda} v_{\phi} \sim 10^{-5}$~eV. 

Quantum consistency requires $\lambda_{\phi}, \lambda_S \gsim \lambda^2 / 16 \pi^2$, $\lambda_{\phi} \gsim y_{\psi, n}^4 / 16 \pi^2$, and $\lambda_S \gsim g^4 / 16 \pi^2$; conditions that are easily satisfied for this choice of parameters.  Additionally, after $S$ gets a vev there is an $\mathcal{O}(1)$ correction to $m_{\phi}$ as well as a small mixing between $\phi$ and $S$. However none of these effects change our discussion in any significant way.

Any $\Pi$s that do not annihilate through the above process would decay into $n \bar n$ with a lifetime
\beq
\tau (\Pi\to n\bar n) \sim \frac{16 \pi}{m_\Pi} \left(\frac{v_\phi^2 m_\phi^2}{\mu^2 m_\psi \, m_n}\right)^2 \frac{1}{\sin^2(\theta_\mu^{eff})} \,,
\label{tauPi}
\eeq
where, the effective CP violating angle $\theta_\mu^{eff} \sim a/f_a$ is expected to be $\sim 1$ initially.  Using the same parameters as above, we find $\tau (\Pi \to n \bar n) \sim 10^{18} \csc^2(\theta_\mu^{eff})$~eV$^{-1}$.  Note that the Hubble time at 
$T\sim 100$~eV, near the onset of FDM oscillations, is $\sim 10$~yr.  Hence, for $m_a \sim 10^{-22}$~eV $\sim 2 \pi /$~yr, the angle
$\theta_\mu^{eff} \sim 1$  does not get redshifted appreciably over many cycles.  
We then expect that the $\Pi$ population will decay efficiently, with a 
lifetime of $\sim 10$~min, into $n \bar n$ and turn into $\mu$-sector radiation,  
long before the standard era of matter domination corresponding to $\sim 10^5$~yr.

In this Letter we demonstrated that a ``fuzzy Dark Matter'' axion can obtain its mass of $\ord{10^{-22}~\text{eV}}$ from infrared confining dynamics at a scale of $\ord{10^2~\text{eV}}$ while being consistent with various cosmological constraints. The scenario we proposed typically leads to a deviation from the standard cosmology 
prediction for the effective number of neutrinos, and is potentially testable 
in neutrino oscillation experiments that search for sterile neutrinos.

\acknowledgments
We thank Michael Creutz and Robert Pisarski for useful discussions.
This work is supported by the United States Department of Energy under Grant Contract DE-SC0012704.


%

\end{document}